# Non-bulk Superconductivity in $Pr_4Ni_3O_{10}$ Single Crystals Under Pressure

Xinglong Chen[1,*], E.H.T. Poldi[2,3], Shuyuan Huyan[4,5], R. Chapai[1], H. Zheng[1], S. L. Bud'ko[4,5], U. Welp[1], P. C. Canfield[4,5], Russell J. Hemley[2,6,7], J. F. Mitchell[1] and D. Phelan[1,*]

[1]*Materials Science Division, Argonne National Laboratory, Lemont, IL 60439, USA*

[2]*Department of Physics, University of Illinois Chicago, Chicago, IL 60607, USA*

[3]*Advanced Photon Source, Argonne National Laboratory, Lemont, IL, USA*

[4]*Ames National Laboratory, Iowa State University, Ames, IA 50011, USA*

[5]*Department of Physics and Astronomy, Iowa State University, Ames, IA 50011, USA*

[6]*Department of Chemistry, University of Illinois Chicago, Chicago IL 60607*

[7]*Department of Earth and Environmental Sciences, University of Illinois Chicago, Chicago IL 60607*

* Corresponding Authors: dphelan@anl.gov *(D.P.)*, xinglong.chen@outlook.com *(X.C.)*

## Abstract

Magneto-transport measurements of $Pr_4Ni_3O_{10}$ single crystals performed under externally applied pressures up to 73 GPa in diamond anvil cells with either KBr or Nujol oil as pressure media yield signatures of superconductivity with a maximum onset temperature of approximately 31 K. True zero resistance was not observed, consistent with a non-percolating superconducting volume fraction. Magnetization measurements provided corroborating evidence of superconductivity, with a pressure-dependent diamagnetic signal occurring below the onset temperature, and an estimate from the absolute value of the susceptibility suggests a superconducting volume fraction on the order of 10%. We observe sample-to-sample variations in the magnitude and pressure dependence of $T_c$ as well as a dependence on the configuration of electrical contacts on a given sample. Possible causes of this behavior may be significant inhomogeneities in the pressure and/or damage to the samples induced by the pressure media as well as inhomogeneities in the crystals themselves. The results imply that our as-grown $Pr_4Ni_3O_{10}$ single crystals are not bulk superconductors but that there is a minority structure present within the crystals that is indeed superconducting.

**Introduction**

In 2023, superconductivity was reported for Ruddlesden-Popper (RP) $La_3Ni_2O_7$ crystals under externally applied pressure [1–5]. This was a surprise since most researchers had approached superconductivity in nickelates from the perspective of searching for it in materials with electron counts that matched those of superconducting cuprates, i.e. by looking at materials with $Ni^{1+\delta}$ cations analogous in electron count to $Cu^{2+\delta}$ [6–8]. This latter approach had indeed yielded superconductivity in thin films of the so-called 112 infinite layer, square-planar nickelates such as $Nd_{1-x}Sr_xNiO_2$ [9–11] and in other square-planar nickelate thin films such as $Nd_6Ni_5O_{12}$ [12]. Nevertheless, no bulk nickelate superconductor has ever been reported for the square-planar, $Ni^{1+\delta}$ containing materials [13,14]. The report of $La_3Ni_2O_7$ thus generated significant interest because: (1) the superconductivity was observed in bulk crystals; (2) the superconductivity occurred in a compound with Ni in octahedral coordination, having a formal valence of $Ni^{2.5+}$, which is a significant departure from cuprates; and (3) the observed superconducting critical temperature ($T_c$) was as high as 80 K. Follow-up work indicated that two forms of $La_3Ni_2O_7$ exist, one with a classic bilayer structure, and the other with alternating single layer and trilayer structures [1,15,16]. Key questions for $La_3Ni_2O_7$ remain, including whether or not it possesses charge or spin density waves (CDWs or SDWs) under ambient conditions and whether such instabilities are correlated to superconductivity, how the superconductivity (or lack of) is manifest in the two structural polymorphs, whether the superconductivity is truly a bulk phenomenon or a filamentary effect [3,17], how the superconductivity is affected by structural defects that are prevalent in $La_3Ni_2O_7$ [18], and how superconductivity is impacted by stoichiometry [18] such as oxygen control.

Subsequent to the report of $La_3Ni_2O_7$, evidence for superconductivity was reported by several groups for RP $La_4Ni_3O_{10}$ under pressure [17,19,20]. Not only did these new reports show that $La_3Ni_2O_7$ was not just a one-off, but they were also important because $La_4Ni_3O_{10}$ specimens are much easier to grow as large crystals in a floating zone furnace than $La_3Ni_2O_7$. This is because $La_4Ni_3O_{10}$ is stable under a fairly broad range of oxygen pressures in contrast to $La_3Ni_2O_7$ [21]. This narrow stability window for $La_3Ni_2O_7$ leads to small crystals and significant inhomogeneity. Moreover, in $La_4Ni_3O_{10}$ the formal Ni valence is $Ni^{2.67+}$ instead of $Ni^{2.5+}$, indicating that the superconductivity is stable across a range of nickel charge concentrations.

$La_4Ni_3O_{10}$ has been more thoroughly studied than $La_3Ni_2O_7$ under ambient pressure. Indeed, it is already established that $La_4Ni_3O_{10}$ develops a CDW at $T_{CDW} \approx 148$ K [22], which is evidenced by a metal-metal transition in the temperature-dependent electrical transport with a kink occurring at $T_{CDW}$, a peak in the heat capacity at $T_{CDW}$, and a concomitant step in the magnetic susceptibility. Single crystal X-ray diffraction experiments evidenced a sinusoidal CDW on all three layers constituting a trilayer [22], and single crystal neutron diffraction evidenced a quasi-2D SDW coupled to the CDW occurring on the top and bottom layers of a trilayer with a node occurring on the middle layer [22]. Pressure-dependent resistance measurements on $La_4Ni_3O_{10}$ revealed a disappearing CDW, which suggests that superconductivity may be correlated to the suppression of a CDW with pressure [17]. A complicating factor for $La_4Ni_3O_{10}$ research is that it possesses two polymorphs that can be present after floating zone growth, one monoclinic and the other orthorhombic [21]. While both polymorphs possess the trilayer RP structure, they do have different bonding characteristics and geometries given their different symmetries, and it is unclear at this point if one or both polymorphs superconduct.

To explore the generality of pressure-induced superconductivity in RP nickelates, we carried out pressure-dependent transport measurements on another RP trilayer nickelate crystal, $Pr_4Ni_3O_{10}$. Similar to $La_4Ni_3O_{10}$, $Pr_4Ni_3O_{10}$ has been characterized thoroughly under ambient pressure. It also develops a coupled CDW and SDW [21,23] at $T_{CDW}\approx158$ K. Nevertheless, there are some potentially important differences that could affect pressure-induced superconductivity. First, $Pr^{3+}$ cations can be magnetic or possess a nonmagnetic singlet state. At ambient pressure, the exchange field from the Ni SDW induces moments on $Pr^{3+}$, which develop a SDW that is coupled to the SDW on nickel sites [23]. How the $Pr^{3+}$ magnetism is affected by pressure or superconductivity is not known. Second, $Pr^{3+}$ cations have a smaller ionic radius than $La^{3+}$, leading to a smaller lattice volume for $Pr_4Ni_3O_{10}$ compared to $La_4Ni_3O_{10}$. Thus, there is a chemical pressure that might affect the external pressures needed to achieve superconductivity. Third, unlike the case of $La_4Ni_3O_{10}$, there is only a single monoclinic phase of $Pr_4Ni_3O_{10}$ that is stable, simplifying the interpretation of the behavior. Finally, $Pr^{3+}$ cations are well-known to participate in charge transfer via hybridization with O in some compounds, a prominent example being the "Pr anomaly" in cuprates [24–26] where the hybridization between Pr and O orbitals suppresses superconductivity in $PrBa_2Cu_3O_7$. $Pr_4Ni_3O_{10}$ allows us to explore what, if any, role such hybridization might play in the superconducting nickelates.

In this report, we present the results of resistance and magnetization measurements on $Pr_4Ni_3O_{10}$ crystals performed in diamond anvil cells (DACs) using different pressure transmitting media. The signature of the CDW, which appears as an anomaly at approximately zero pressure in the resistance, was observed to disappear under pressure. The signature of superconductivity appears as a downturn in the resistance on decreasing temperature, visible under pressures exceeding a rather low threshold of $P\approx10$ GPa, and resistance drops up to ~98% were observed. Although zero resistance was not achieved in the attainable temperature range, the similarity of the results to those reported for $La_4Ni_3O_{10}$ [17], the systematic evolution with pressure, and the variation of the transition with applied magnetic field argue that we are indeed probing superconductivity. Variability among samples and/or current path dependence was observed, which we attribute to inhomogeneous pressure and intrinsic sample defects. Pressure and temperature dependent diamagnetism was also observed, corroborating the observation of superconductivity. However, the low (~10%) superconducting volume fraction indicates that we are observing non-percolating superconducting fragments that possess a structure that is *not bulk stoichiometric* $Pr_4Ni_3O_{10}$.

**Methods**

Single crystals of $Pr_4Ni_3O_{10}$ were grown at 140 bar $O_2$ pressure using a SciDre HKZ floating zone furnace at Argonne National Laboratory. The growth was previously described in detail in Ref. [21], and extensive characterization of the magnetic phase behavior of our crystals under ambient pressure can be found in our recent report [23].

For the high-pressure resistance experiments, we employed two different kinds of pressure transmitting media (PTM), solid KBr and Nujol oil. A more thorough description of the measurements with KBr PTM is provided in **SI Section I**, but succinctly the contacts were located along the periphery of a representative sample in a van der Pauw geometry, though the van der Pauw conditions (e.g., contact size << sample dimensions; uniformity of resistance) were clearly not met, as described below. We thus report measured resistances rather than resistivities. The four contacts on the periphery allow for two separate four terminal resistance pairs to be measured in which the two current and two voltage contacts are roughly parallel to one another. We will refer to these two configurations as $R_1$ and $R_2$. We measured two samples of $Pr_4Ni_3O_{10}$ using the KBr PTM. The first, which we will refer to as *Sample 1*, consisted of one small crystal in the pressure cell, and its two resistances are denoted $R_1^{S1}$ and $R_2^{S1}$. The second, *Sample 2*, consisted of several small crystals placed together into the pressure cell, and its two resistances are denoted $R_1^{S2}$ and $R_2^{S2}$. For the resistance measurements with the Nujol oil, we also present results for two samples, (noted *Sample 3* and *Sample 4*). A Be-Cu DAC (Bjscistar) that fits into a Quantum Design PPMS was used with 400 μm (up to 50 GPa) or 350 μm (up to ~73 GPa) culet, standard-cut type Ia diamonds. The sample was cut and polished into a thin flake and loaded together with a tiny ruby sphere into an apertured, stainless-steel gasket covered by cubic-BN. For these measurements, a conventional 4-terminal geometry was employed with the four contacts positioned along the length of the sample in the order of $I^+$, $V^+$, $V^-$, $I^-$ as pictured in Fig. S1(c).

Finally, magnetization measurements were performed with a SQUID magnetometer (Quantum Design MPMS) and the sample (*Sample 5*) in a DAC using 400 μm culet diamonds, a tungsten gasket, Nujol mineral oil as the PTM, a ruby for calibration, Stycast, and a cell body (easyLab Mcell Ultra) made out of Be-Cu alloy (see Fig. S1(d)). Background DC magnetization measurements were performed as a function of increasing temperature in an applied field $\mu_0H$=0.02 T following zero-field cooling warming measurement under with all the components of the cell save for the sample. The SQUID voltage versus sample position traces were collected at each temperature at P~0 GPa. Then the sample was loaded, and the measurements were re-performed as a function of temperature and pressure. The background SQUID voltage traces at each temperature were subtracted from the corresponding traces with the sample, and the resultant voltage versus position traces were fit to that expected for a dipole to extract the sample moment. Additionally, flux trapping measurements were performed by cooling the DAC from 50 K to 2 K in $\mu_0H$=2 T, stabilizing the temperature at 2 K for 10 minutes, and then removing the field. The magnetization was then measured upon warming from 2 K to 45 K.

As stated above, pressure determination was performed using pieces of ruby crystals as calibrants. In these measurements, the fluorescence of the ruby is measured at ambient temperature. This emission becomes weaker and broader as the pressure is increased, leading to uncertainty or imprecision in the determination of pressure. Moreover, imperfect hydrostatic conditions can lead to non-uniform pressures on the sample itself as well as differences between the measured ruby pressures and the sample. We also noticed differences amounting to up to several GPa in measured ruby pressures before and after experiments using the KBr PTM, indicative of irreversible changes in pressure during experiments due to thermal contraction and expansion. Given all these factors, the pressures listed here are considered as estimates with uncertainties of order 10%. Such uncertainties have no significant consequence at the level of analysis presented here.

**Results**

*Electrical Transport with KBr PTM*

Figs. 1(a) and 1(b) show the resistances, $R_1^{S1}$ and $R_2^{S1}$, for *Sample 1* as a function of temperature (*T*) at different pressures (*P*). Figs. 2(c) and 2(d) display the same data over a different ordinate axis scaling. Of immediate note is that at ambient pressure, both $R_1^{S1}$ and $R_2^{S1}$ show anomalies at $T_{CDW}$; however, they exhibit markedly different temperature dependences. While both show metallic behavior ($dR/dT > 0$) above $T_{CDW}$, $R_1^{S1}$ increases by approximately a factor of three below $T_{CDW}$ in almost a step-like fashion, whereas $R_2^{S1}$ has a small cusp at $T_{CDW}$, a return to metallic behavior upon further decrease in temperature, and then an insulating upturn at the lowest temperatures. These differences highlight the inhomogeneous conduction through the sample, with the measured resistance strongly depending on the current path. We note that a very similar phenomenology was reported on $La_4Ni_3O_{10}$ [17]. An increase in the pressure to *P*~3.8 GPa suppresses the magnitude of the stepwise change in resistance at $T_{CDW}$ in $R_1^{S1}$, and an evolution of the cusp in $R_2^{S1}$ to a slightly insulating upturn upon decreasing temperature is observed. For both permutations, the resistance decreases with the increased pressure throughout the temperature range.

$R_1$ and $R_2$ behave differently near 10 GPa. With decreasing temperature, metallic behavior is observed in $R_1^{S1}$ down to about 45 K, below which a slightly insulating upturn is observed. A cross-over to lower resistance is observed below $T_c$ of ~17 K. However, the absence of *R*=0 implies that the putative superconducting regions do not form a connected path between the voltage contacts. The current path dependence is even more pronounced in the case of $R_2^{S1}$. Indeed, there is no signature of superconductivity in $R_2^{S1}$. Increasing pressure to ~16 GPa enhances $T_c$ in $R_1^{S1}$, with a concomitant drop in the magnitude of the resistance; however, there is still no signature of a superconducting transition in $R_2^{S1}$. For pressures in the range of ~ 25 to 55 GPa, the superconducting transition is clearly visible in $R_1^{S1}$ and also $R_2^{S1}$, where a local maximum is observed in the resistance at T ≈ $T_c$ which was observed in $R_1^{S1}$. At even higher pressures, the decrease in $R_1^{S1}$ becomes less sharp and less pronounced, and the signature of superconductivity again disappears from $R_2^{S1}$. Thus, pressures above ~ 40 GPa suppress the superconductivity in *Sample 1*, implying the existence of a dome in $T_c(P)$, as described further below.

We measured the temperature dependence of $R_1^{S1}$ at *P* ~32 and ~55 GPa under magnetic fields up to $\mu_0H$=9 T, as shown in Figs. 2a and 2b. Under increasing field, we observe the systematic shift of the transition to lower temperatures, as would be expected for a superconductor. Furthermore, the shift occurs in an essentially parallel fashion without any field-induced broadening, suggesting that effects due to fluctuations and/or granularity are not significant. Using the 90% normal-state resistance as the criterion for $T_c$, we trace the upper critical field $H_{c2}$ as shown in Fig. 3c for 32 and 55 GPa. At $P\approx32$ GPa, $H_{c2}$ increases rapidly with decreasing temperature, with an average slope of $\mu_o dH_{c2}/dT|_{T_c}$=-2.5 T/K corresponding to a short Ginzburg-Landau (GL) coherence length of $\xi_{GL}\approx2.5$ nm. We estimate $H_{c2}(0\,K)$ of 44 T at 32 GPa and 16 T at 55 GPa using phenomenological fits according to $H_{c2}(T) = H_{c2}(0)\,(1 - t^2)/(1 + t^2)$ with $t = T/T_c$ [27,28] as shown in Fig. 2c. Our estimate of $H_{c2}$ is nearly the same as that reported for La$_4$Ni$_3$O$_{10}$ with similar $T_c$ (at 69 GPa) [17]. While these values are close to the BCS paramagnetic limit of $\mu_0 H_P[T] = 1.8\,T_c[K]$ [29,30], we note that the estimated upper critical fields are sensitive to the criteria employed to define $T_c$. Illustrating this, we have also included phenomenological fits to the same formula using the onset to the transition as the definition of $T_c$. In contrast, the upper critical field slopes are much less sensitive to the criterion, as the transitions shift in a parallel fashion.

Transport data were also collected for *Sample 2* under the KBr PTM as shown in SI Fig. 2 and discussed in more detail in the **SI Section II**. The signature of superconductivity in *Sample 2* was significantly weaker and manifest as a field-dependent kink with a much smaller relative change in resistance, consistent with an onset of trace superconductivity.

*Electrical Transport with Nujol as PTM*

We also performed resistance measurements with Nujol mineral oil as the PTM. The Nujol oil is also an imperfect solid PTM in the *P-T* phase space of interest but its shear strength on compression is expected to differ from that of KBr. Fig. 3(a,b) show the temperature dependence of the resistance of *Sample 3* as a function of pressure. In this case, we observed an increase in the resistance upon cooling below $T_{CDW}$ near ambient pressure. The magnitude and onset temperature of this increase systematically diminished as the pressure increased, but the CDW was still apparent at least up to ~20 GPa. This is a clear point of difference from the results found from KBr PTM, where the CDW has disappeared at 10 GPa [see Fig. 2(a)] and is surprising given the relatively low pressures required to suppress the CDW in La$_4$Ni$_3$O$_{10}$, as to a first approximation one would expect the CDW would have a similar stability in Pr$_4$Ni$_3$O$_{10}$ given its similar $T_{CDW}$. We would thus suggest that future work using a Be-Cu piston cell at lower pressures may be required to establish the precise pressure dependence of the suppression of the density wave.

We observed sharp decreases in the resistance of *Sample 3* at approximately 37 GPa and 42 GPa at ~13 K, characteristic of a superconducting transition again in a minority, non-percolating volume fraction of the sample with an onset $T_c$ of ~13 K (Fig. 3(a,b)). At the maximum pressure ($P\approx48$ GPa), a slight decrease in $T_c$ was observed along with a smearing of the transition, suggesting that 13 K might have been the maximum $T_c$ for this sample. A possible explanation of this behavior is that the stoichiometry of superconducting fragments in this sample differs from that of *Sample 1*.

The pressure-dependent electrical transport was also measured for *Sample 4* with Nujol mineral oil as the PTM, as shown in Fig. 3(c,d). Again, we observed the CDW at low pressures and a suppression with increasing pressure. In this case, a remnant of the CDW was apparent at $P\approx16$ GPa, whereas at ~24 GPa and above, metallic behavior was observed. Starting near 30 GPa, we observed a low temperature decrease in the resistance, again signaling superconductivity, evolving from an onset $T_c$ of 12 K at 30 GPa to maximal $T_c$ of 23 K at 49 GPa. Importantly, a 98% change in resistance was observed with the resistance becoming very close, but not quite zero in the attainable temperature range. Measurements performed with the $I^+$ and $V^+$ leads transposed evidence a current path dependence on the change in resistance which is manifest due to the inhomogeneous transport as described in **SI Section III**.

Combining the results from the electrical transport under pressure from all four samples, we plot the observed $T_c$ as a function of pressure in Fig. 4a. Here we define $T_c$ by the onset temperature of the downturn in resistance. It is immediately apparent that there is considerable sample to sample variability in both $T_c$ and the evolution of $T_c$ under pressure. *Sample 1* possessed an obvious a broad-topped dome with a maximum $T_c \approx 31$ K at $P\approx45$ GPa. However, *Samples 3* and *4* have lower $T_c$'s with an onset of superconductivity occurring at higher pressures. It is thus clear that the superconducting fragments in these samples are not a "line phase" but rather express that a currently unknown variable (e.g., stoichiometry, strain, bonding, etc.) affects the $T_c$ as well as the position of the superconducting dome in *P-T* parameter space. We also plot the percent change in the resistance, $\Delta_R=100\% \times [R_1^{S1}(T_c)-R_1^{S1}(2\text{ K})]/R_1^{S1}(T_c)$ as a function of pressure in Fig. 4b. Again, this yields dome-like behavior, but the sample-to-sample variability in the magnitude of $\Delta_R$ as well as the position of the dome is even more obvious.

*Magnetic Susceptibility*

*T*-dependent DC magnetic susceptibility ($\chi$) data under $P\approx5$, 16, 35, and 48 GPa, with the empty cell measurements subtracted out as a background, are shown in Fig. 2(d). At $T\approx40$ K, the magnetization at all *P* converges to $\chi \approx 0$, which is the baseline. However, below 30 K, the traces for the different pressures diverge. Whereas at $P\approx5$ GPa, a weak Curie tail with positive $\chi$ is observed, a diamagnetic signal is observed and enhanced as *T* is lowered below $T\approx30$ K for the higher *P*. This diamagnetic signal is enhanced with increasing *P*. Given that the data diverge around the maximum onset $T_c$ measured in the resistance measurements of *Sample 1* under pressure, a reasonable interpretation is that the diamagnetism is a magnetic field screening that has an onset at the maximum $T_c$ regardless of pressure because of the pressure inhomogeneities. Based on the approximate dimensions of the sample and a calculated demagnetization factor, the magnitude of the diamagnetic signal yields a superconducting volume fraction of 6% at 10 K and 48 GPa, assuming that the diamagnetic signal is entirely due to a superconducting screening. Note that we regard these volume fractions only as an estimate since factors such as non-ideal demagnetization geometry can induce large uncertainties. In particular, as *T* is decreased below 10 K, the measured susceptibility continues to decrease, suggesting that the superconducting volume fraction likewise increases; however, the uncertainty in this region becomes unacceptably large due to the very low signal to background ratio, and we have thus designated this temperature regime as questionable in Fig. 2(d). Trapped flux measurements (See **SI Section IV**) yielded negative results, consistent with filamentary or non-bulk superconducting regions that are too small to effectively pin magnetic flux.

**Discussion and Summary**

The magnetotransport and susceptibility data presented here provide compelling evidence for superconductivity in a low volume fraction of order 10%, or less, that does not yield percolating pathways in bulk $Pr_4Ni_3O_{10}$ crystals under pressure. We observe a maximum (onset) $T_c$ of approximately 30 K, which is similar to that reported for $La_4Ni_3O_{10}$ [17]. The data presented here with KBr as the PTM are reminiscent to that reported with KBr for $La_4Ni_3O_{10}$ [17]. Indeed, in neither case is a zero-resistance state found with KBr. Measurements of our crystals with Nujol as the PTM come very close to zero resistance. In the case of $La_4Ni_3O_{10}$, zero resistance was achieved with a helium PTM [17]. Recently, Zhu *et al*. [17] claim 80% superconducting volume fraction in $La_4Ni_3O_{10}$ measured in a field of 20 Oe, and this is significantly higher than what we have estimated for $Pr_4Ni_3O_{10}$. The estimates of upper critical magnetic fields are likewise similar to those published for $La_4Ni_3O_{10}$ and $La_3Ni_2O_7$; thus, it is reasonable to suspect that the mechanism of superconductivity under pressure is the same in all these compounds. Nevertheless, our measurements here unambiguously indicate that our as-grown $Pr_4Ni_3O_{10}$ crystals are *not* bulk superconductors, and it is puzzling why $La_4Ni_3O_{10}$ has essentially the same $T_c$ and $H_{c2}$ as $Pr_4Ni_3O_{10}$ but a much higher superconducting volume fraction.

We address four possible explanations to the above quandary. *Explanation 1: Superconductivity is truly a property of the trilayer RP phase, but it is very sensitive to stoichiometry*. This explanation assumes that our as-grown crystals have a spatially-varying stoichiometry and that only a low volume fraction of our crystals assumes the stoichiometry that superconducts. We note that thermogravimetric measurements performed on our as-grown crystals indicated that their oxygen stoichiometry was very close to nominal with an oxygen content of 10.05(1). This would then imply that the superconductivity is a property of non-stoichiometric $Pr_4Ni_3O_{10}$. The sample-to-sample variation would be understood by some crystals possessing higher levels of the superconducting stoichiometry. Tunability in the stoichiometry would naturally explain the variation in $T_c$. Regions possessing the superconducting stoichiometry could in principle be few and disconnected. *Explanation 2: Superconductivity in our crystals arises from a defect rather than being a property of the pure trilayer RP phase*. RP phases are well-known to be subject to stacking-fault defects or stacking mis-layering, i.e. possessing small regions of RP layers of different layering (in the present case, *n*=2, *n*=4, etc.). These layering mistakes present an opportunity for boundaries at the defects, and it is possible that these superconduct. This is precisely what Zhou *et al*. argue in their study of $La_3Ni_2O_7$ [3]. *Explanation 3: Superconductivity in our crystals is due to an unknown oxide impurity completely unrelated to the RP phase*. We reject this explanation as there is no evidence of a large enough impurity level in our crystals [21] to yield such a significant superconducting volume fraction. Rather the characterization of our crystals suggests the opposite. *Explanation 4: Superconductivity in $Pr_4Ni_3O_{10}$ is extremely* *sensitive to strain, and the non-hydrostatic conditions in the experiment lead to a strain field that yields a low volume fraction of superconductivity.* The numerous materials that exhibit superconductivity via the established techniques employed here suggest that the superconductivity in $Pr_4Ni_3O_{10}$ would be unusually sensitive to strain for this explanation to hold water; nevertheless, this needs to be considered.

The fact that the maximum $T_c$ of $Pr_4Ni_3O_{10}$ observed here is similar to that observed in $La_4Ni_3O_{10}$ [17] suggests, not unsurprisingly, that the physics is dominated by the Ni-O trilayers in these compounds. As alluded to above, oxides can be affected by hybridization of Pr and O orbitals, which can lead to anomalous transport properties in Pr-containing transition metal oxides vis-a-vis similar compounds that possess other rare earths (*e.g.*, La, Nd, etc.). In $Pr_{0.5}Ca_{0.5}CoO_3$, Pr-O hybridization leads to a 1$^{st}$ order metal-insulator transition [31]. In the Pr-123 cuprate, such Pr-O hybridization is understood to suppress superconductivity entirely [24–26]. In the present case, we speculate that the Pr-O hybridization plays a lesser role in nickelates than cobaltites or cuprates. For example, the perovskite $PrNiO_3$ exhibits transport behavior that falls in line with all other rare earth perovskite nickelates except $LaNiO_3$ [32]. In the hole-doped 112 $PrNiO_2$ thin films, superconductivity was also not suppressed by Pr-O hybridization [33]. Finally, whatever the source of the superconductivity is here, it clearly involves all three elements (Pr, Ni, and O) suggesting that Pr-O hybridization is likewise insignificant.

Another unique aspect of $Pr_4Ni_3O_{10}$ in comparison to $La_4Ni_3O_{10}$ is that the $Pr^{3+}$ cations develop a SDW coupled to the Ni SDW [23]. This effect leads to an exchange pathway that couples ordered Ni moments between trilayers, which is not present in $La_4Ni_3O_{10}$. We postulate that since pressure suppresses the CDW, it also suppresses the SDW on the Ni sites, since the CDW and SDW are strongly coupled to one another. The suppression of the Ni SDW naturally suppresses the exchange field felt by the $Pr^{3+}$ sites, so that, $Pr^{3+}$ in $Pr_4Ni_3O_{10}$ under pressure may adopt the expected non-magnetic singlet ground state, akin to that found in perovskite $PrNiO_3$ [34]. Experiments to explore the magnetic behavior of Pr under pressure are planned.

Very recently, Huang *et al.* [35] reported results of pressure-dependent resistance measurements on polycrystalline $Pr_4Ni_3O_{10-\delta}$. They observed a drop in resistance of order 10% upon cooling below ~30 K in zero field at ~ 42 and 54 GPa followed by an upturn upon further cooling below ~ 5 K. Their results were interpreted as a signature of superconductivity. The critical temperature estimated in their work is in agreement with the maximum $T_c$ that we report here, though the estimated values of $H_{c2}(0)$ and the magnitude of the drop in resistance at $T_c$ are significantly lower in the polycrystalline work as might be expected. In the absence of magnetic susceptibility measurements, the superconducting volume fraction of the polycrystalline samples is unknown.

As discussed above, it is likely that both the PTM as well as layering faults contribute to the inhomogeneous electronic conduction observed in our experiments, since there is no perfect hydrostatic PTM at these temperatures. The current paths in the samples strongly depend upon the positions of leads. Thus, different samples show different properties under even ambient pressure. Some samples tend to show more metallic transport than others, and clearly the anomalies that occur in resistance as the CDW sets in are sample dependent. Sometimes this appears as a metal-metal transition with a kink occurring at $T_{CDW}$, whereas other times it appears to be more of a metal-semiconductor transition setting at $T_{CDW}$. Further experiments are needed to determine if there is a strong correlation between pressure-induced superconductivity and the transport characteristics measured at ambient pressure.

Despite the uncertainty in the reason why only a low volume fraction of superconductivity is observed in our measurements, it is also quite clear that $Pr_4Ni_3O_{10}$ is a fertile platform for investigating the relationship between CDWs, SDWs, charge concentrations, and superconductivity. It will be very important to establish the effects of strain, defects, and stoichiometry in follow-up work to determine the origin of the superconductivity. Whatever the explanation, the measurements clearly evidence that a tantalizing superconducting phase is present in these samples. A concerted effort that addresses both the chemistry and materials science (e.g., interplay between stoichiometries, defects, pressure media) of superconducting nickelates on an equal footing with the study of their physics is called for.

**Acknowledgements:**

We thank Alexander Mark for useful discussions and assistance preparing a high pressure cell for resistance measurements. Work at Argonne National Laboratory (crystal growth, resistivity measurements with the KBr pressure medium) was sponsored by the US Department of Energy, Office of Science, Basic Energy Sciences, Materials Science and Engineering Division. Work at UIC was supported by the U.S. National Science Foundation (NSF, DMR-210488), and the U.S. Department of Energy-National Nuclear Security Administration (DOE-NNSA) through the Chicago/DOE Alliance Center (DE-NA0003975), and the DOE Office of Science (DE-SC0020340). Work at Ames National Laboratory (resistance and magnetization measurements with Nujol pressure medium) was supported by the U.S. Department of Energy, Basic Energy Sciences, Material Science and Engineering Division under Contract No. DE-AC02-07CH11358. SLB and PCC acknowledge Mike Vronsky, Steven Pushkov, and Nikanor Chevotarevich for having helped motivate development of trapped flux protocols.

# Figure 1

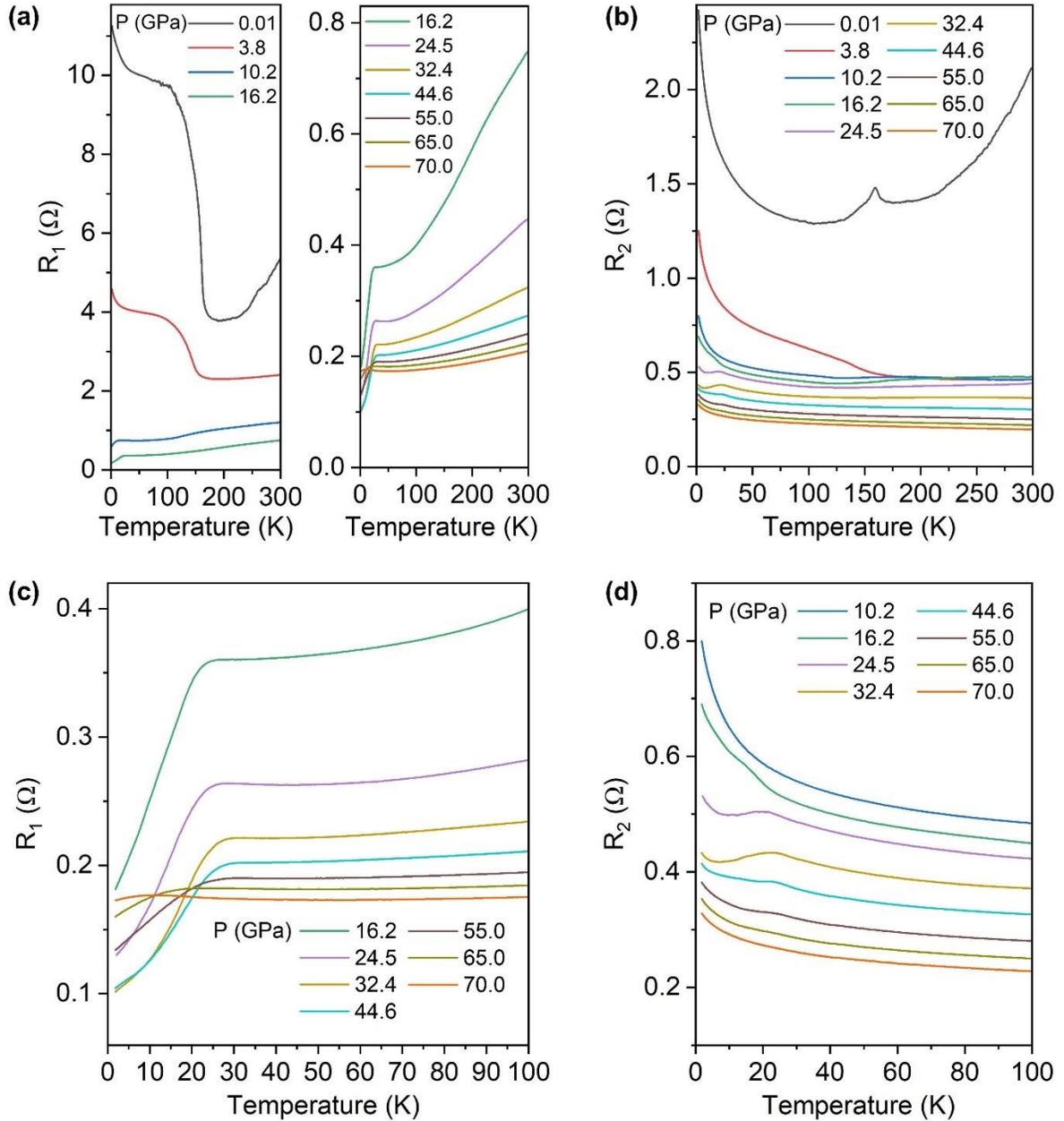

**Figure 2**: Temperature-dependent resistance measurements of *Sample 1* in various pressures as labeled. The PTM was KBr. *$R_1$* and *$R_2$* are shown in (a) and (b), respectively. Panels (c) and (d) display the same data as (a) and (b), respectively, but on a smaller scale to show the low-temperature behavior.

**Figure 2**

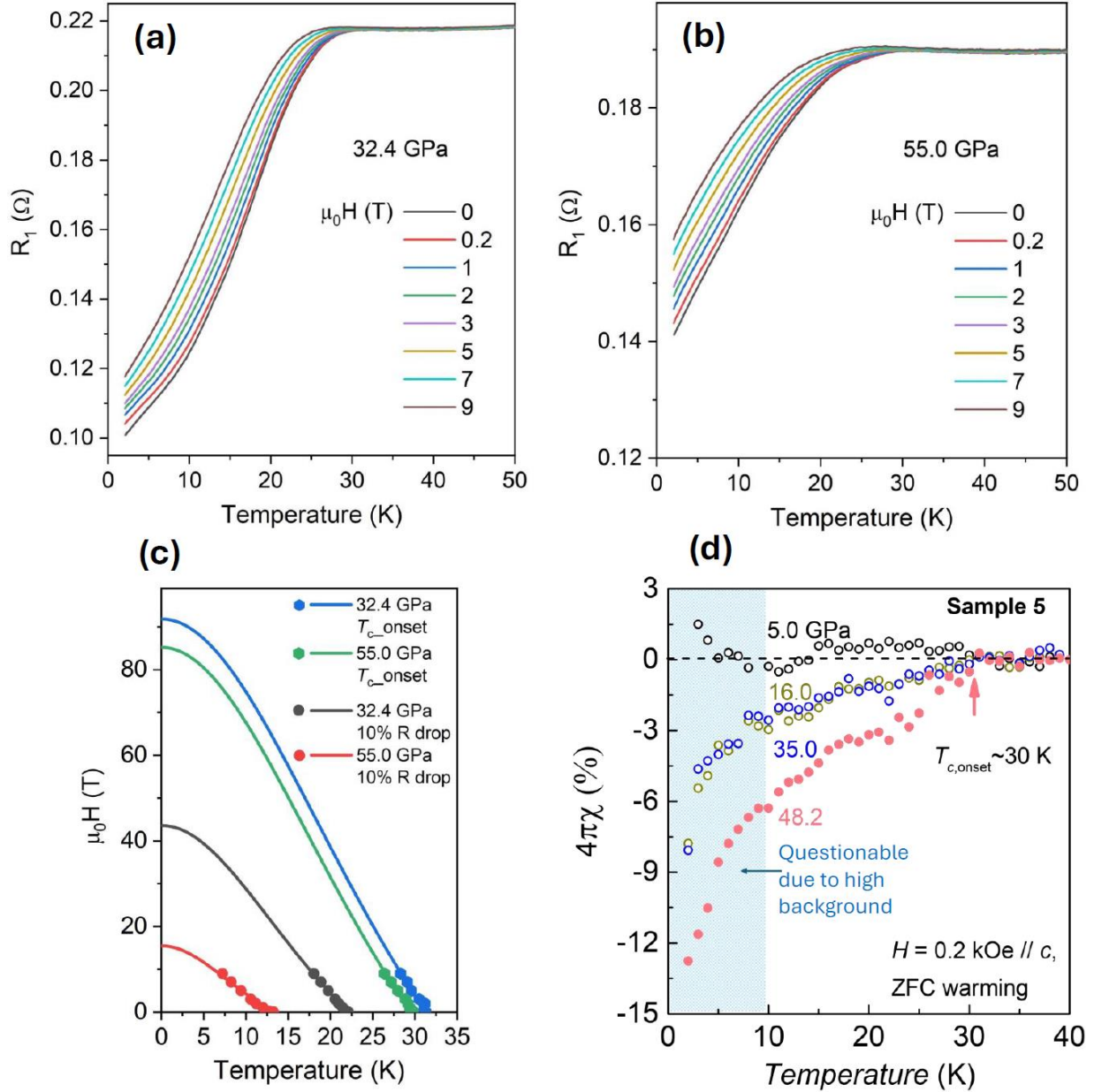

**Figure 2:** Temperature dependent resistance measured in the $R_1$ permutation for *Sample 1* under various magnetic fields at (a) ≈32 GPa and (b) ≈55 GPa. A PTM of KBr was used for the resistance data displayed in (a) and (b). (c) Ginzburg-Landau fits (solid curves) to the extracted $T_C$'s (points) from the various fields. Two definitions of $T_c$ were employed: onset and 10% resistance drop the resistance at onset. (d) Temperature-dependent susceptibility of Sample 5 at various pressures with Nujol as a PTM. As a means of scale, a unit of 1% in $4\pi\chi$ corresponds to approximately $4\times10^{-7}$ emu in magnetization.

# Figure 3

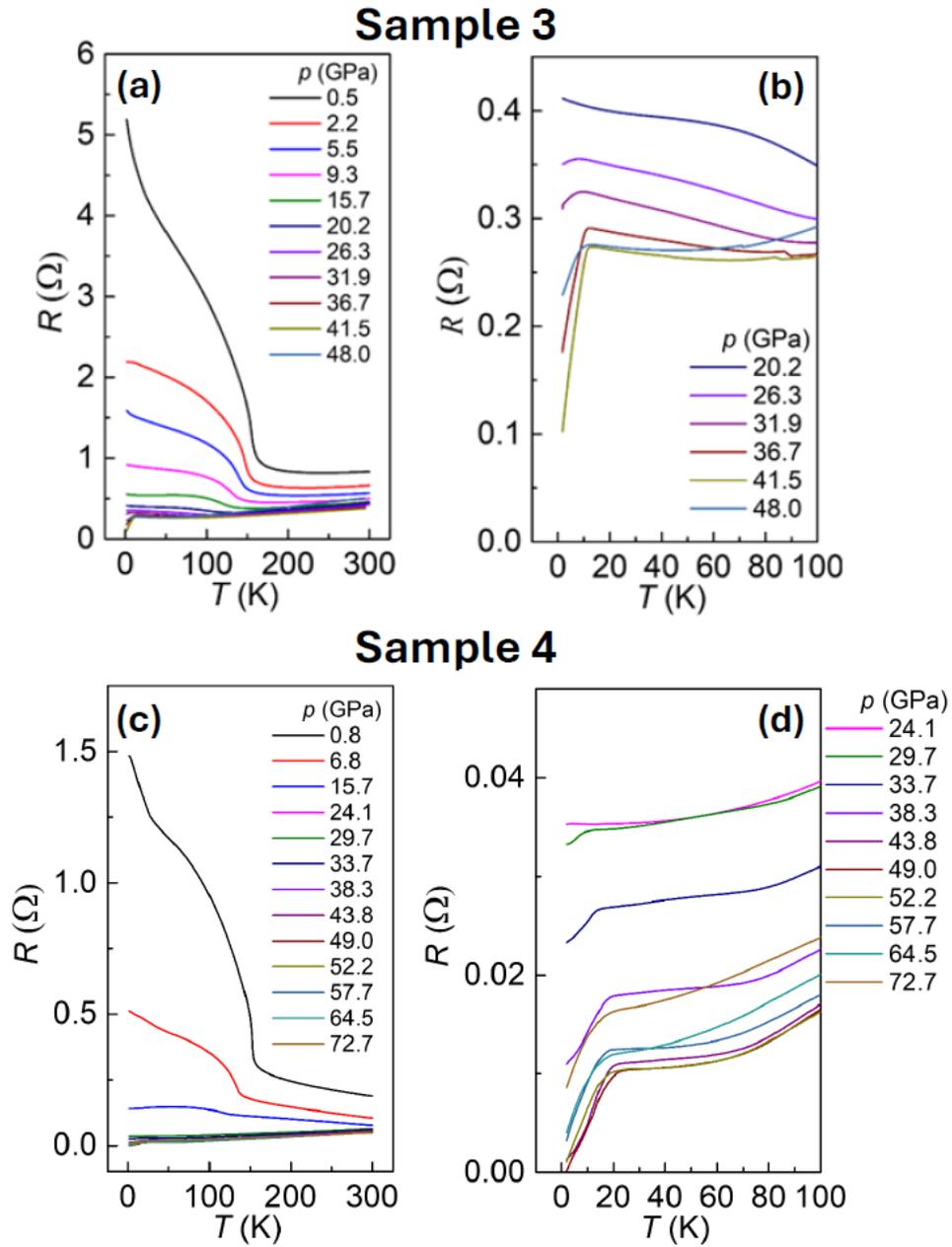

**Figure 3:** (a,b) Temperature-dependent resistance measurements of *Sample 3* measured with the under various pressures as labeled. (c,d) Temperature-dependent resistance measurements of *Sample 4* under various pressures as labeled. Both samples were measured with Nujol as the PTM.

# Figure 4

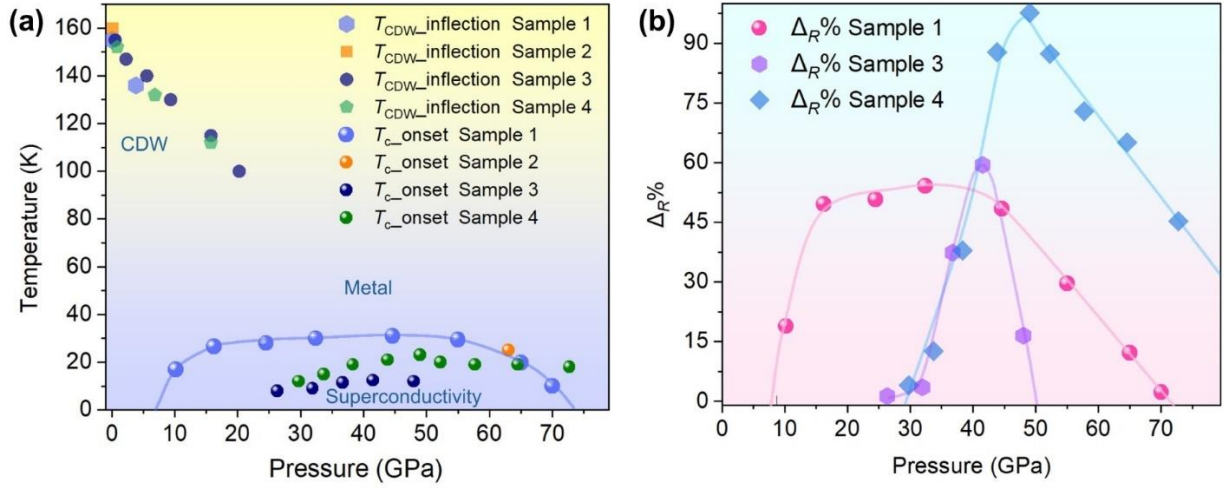

**Figure 4:** (a) Onset $T_c$ and $T_{CDW}$ as a function of pressure. (b) Normalized change in resistance below $T_c$ as a function of pressure. The lines in (a) and (b) are guides to the eye. The data from *Sample 1* and *Sample 2* were collected using a PTM of KBr. The data from *Sample 3* and *Sample 4* were collected using a PTM of Nujol.

Supporting Information for

# Supporting Information for Non-bulk Superconductivity in Pr$_4$Ni$_3$O$_{10}$ Single Crystals Under Pressure

Xinglong Chen[1], E.H.T. Poldi[2,3], Shuyuan Huyan[4,5], R. Chapai[1], H. Zheng[1], S. L. Bud'ko[4,5], U. Welp[1], P. C. Canfield[4,5], R. J. Hemley[2,6,7], J. F. Mitchell[1] and D. Phelan[1]

[1]*Materials Science Division, Argonne National Laboratory, Lemont, IL 60439, USA*

[2]*Department of Physics, University of Illinois Chicago, Chicago, IL 60607, USA*

[3]*Advanced Photon Source, Argonne National Laboratory, Lemont, IL, USA*

[4]*Ames National Laboratory, Iowa State University, Ames, IA 50011, USA*

[5]*Department of Physics and Astronomy, Iowa State University, Ames, IA 50011, USA*

[6]*Department of Chemistry, University of Illinois Chicago, Chicago IL 60607*

[7]*Department of Earth and Environmental Sciences, University of Illinois Chicago, Chicago IL 60607*

# I. Method: Electrical Transport & Diagrams of Cells

For the measurements with the KBr PTM, measurements were carried out in a CuBe pressure cell (Almax EasyLab Diacell CryoDAC-PPMS) using double-beveled 200 µm-culet diamond anvils and CuBe gaskets (see Fig. S1(a,b). After initial gasket indentation to ~10 GPa, a concentric 200 µm-hole was laser drilled into the CuBe where a cured mixture of 10:1 cBN:epoxy (EPO-TEK 353ND) was packed in excess. After compressing this powder mixture to 25-30 GPa, it formed a rigid insert, and the powder excess formed a cBN insulating layer around the gasket, between the CuBe and the diamond side facets. The insert was then laser drilled with a ≈ 100 µm-diameter hole to create a region for adding KBr, which was packed and pressurized to 10 GPa to act as the PTM. The sample was loaded on the KBr PTM together with a ruby ball for pressure calibration. 4 µm-thick Pt foil was used as electrodes, attached on the diamond facets with conducting Ag paint and connected to Cu wires. Extra care was taken to insulate the Pt electrodes from the metallic gasket with a layer of Stycast epoxy.

A picture of the setup for the electrical resistance measurement with Nujol is shown in Fig. S1(c), and a picture of the setup for the magnetic susceptibility measurement is shown in Fig. S1(d).

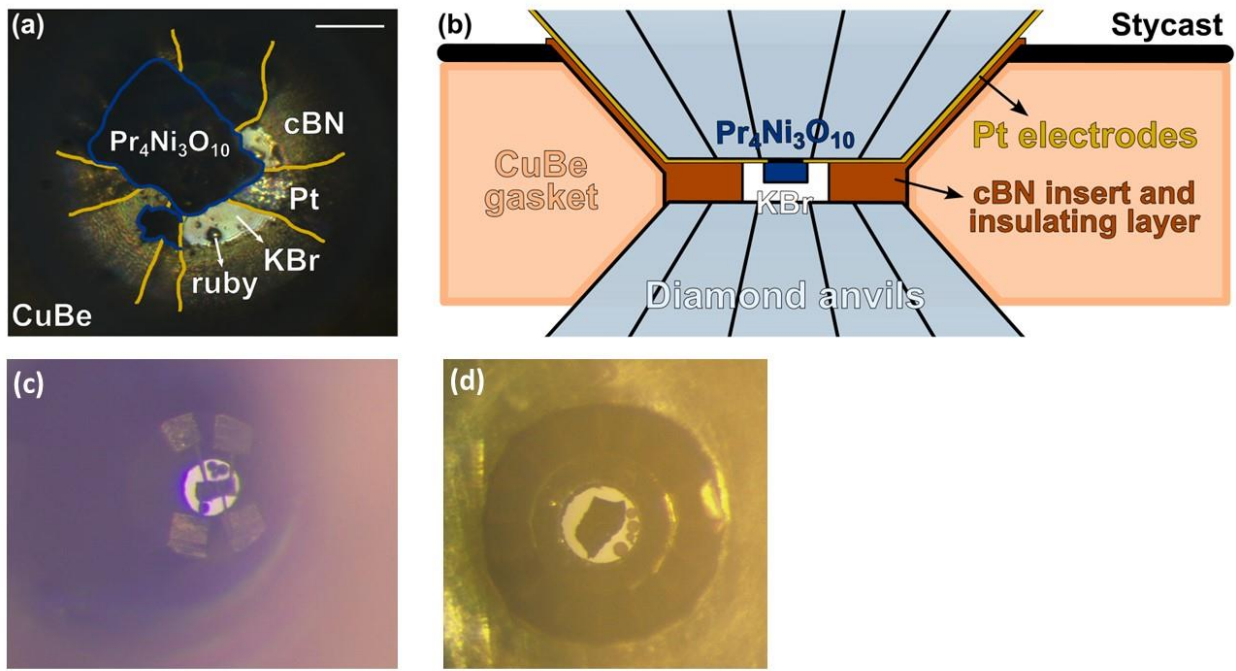

**Figure S1:** (a) Picture of *Sample 1* and electrodes at 70 GPa, seen through the diamonds, using transmitting and reflecting light. The horizontal scale bar represents 50 µm. (b) Schematics of the high-pressure experimental setup for electrical transport under the KBr PTM. (c) Picture of the pressure cell used for electrical transport under the Nujol PTM. (d) Picture of the pressure cell used for magnetization.

## II. Electrical Transport of Sample 2 Under KBr Pressure Medium

Figs. S2(a) and S2(b) show the temperature dependences of the two resistance permutations for *Sample 2*, $R_1^{S2}$ and $R_2^{S2}$ under various pressures. In contrast to *Sample 1*, both permutations show metallic behavior over a wide temperature range at the lowest pressure, with only a slight kink occurring at $T_{CDW}$. The magnitude of $R_2^{S2}$ is much lower than $R_1^{S2}$ lending to considerable noise at low temperatures. In fact, below 75 K, $R_2^{S2}$ reads zero or slightly below zero. We stress that this is *not* due to superconductivity; rather, it is an experimental artifact likely resulting from an unusual path being traversed by the current due to inhomogeneity. As such, $R_2^{S2}$ is not useful for exploring potential pressure-induced superconductivity, but we show it to demonstrate the limitations of the experiment. Focusing on $R_1^{S2}$, we see that as the pressure is increased, $R_1^{S2}$ decreases, and hence the metallic character is enhanced; however, the overall shape of the temperature dependence under pressure is unusual in that it is concave down. Interestingly, at the highest pressures, we observed a kink, in which $R_1^{S2}$ decreases upon further cooling, as shown under magnification in Fig. S2(c). This kink, and its suppression under magnetic field, as shown in Fig. S2(d), may indicate trace superconductivity in *Sample 2* under pressures.

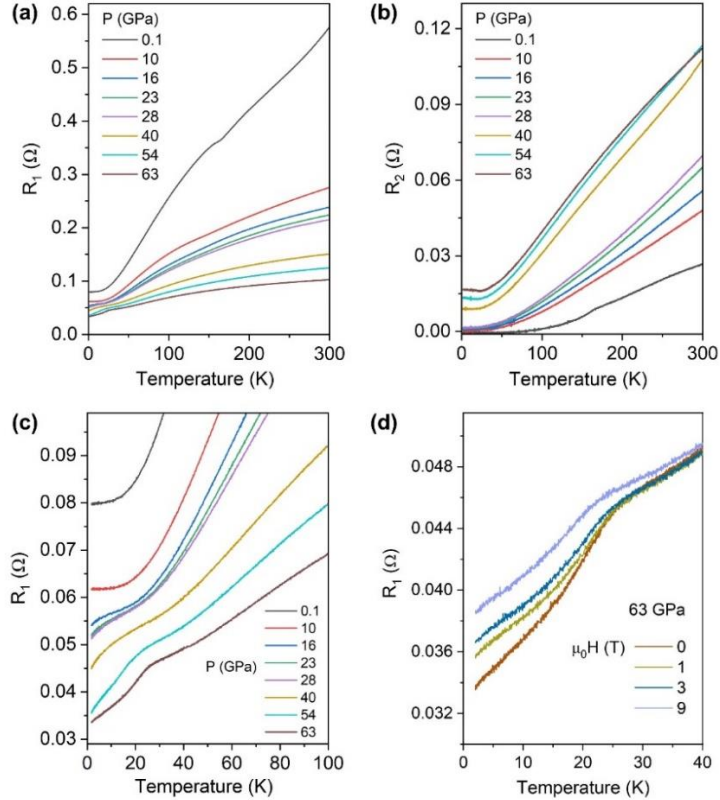

**Figure S2:** Temperature-dependent resistance measurements of *Sample 2* in various pressures as labeled. The PTM was KBr. *$R_1$* and *$R_2$* are shown in (a) and (b), respectively. Panel (c) shows the same data in panel (a) over a narrower range so that the kink can be observed. (d) The field dependence of *$R_1$* at *P*=63 GPa.

## III. Transposed Contacts for Sample 4

In addressing the issue of sample 4's resistance not reaching absolute zero, we applied a non-standard method to test the resistance in slightly different regions, as shown in Fig. S3. By switching the V+ and I+ channels, we found that at 49.0 GPa, the $R(T)$ curve in (b) exhibits a lower onset $T_c$ and a smaller resistance drop ratio, $\Delta R$, compared to the curve in (a). Conversely, at 72 GPa, the $R(T)$ curve in (b) shows a higher onset $T_c$ and a larger $\Delta R$ than in (a). This suggests a delayed response to the applied pressure in the outer region of the sample, possibly revealing the relaxation of a pressure inhomogeneity.

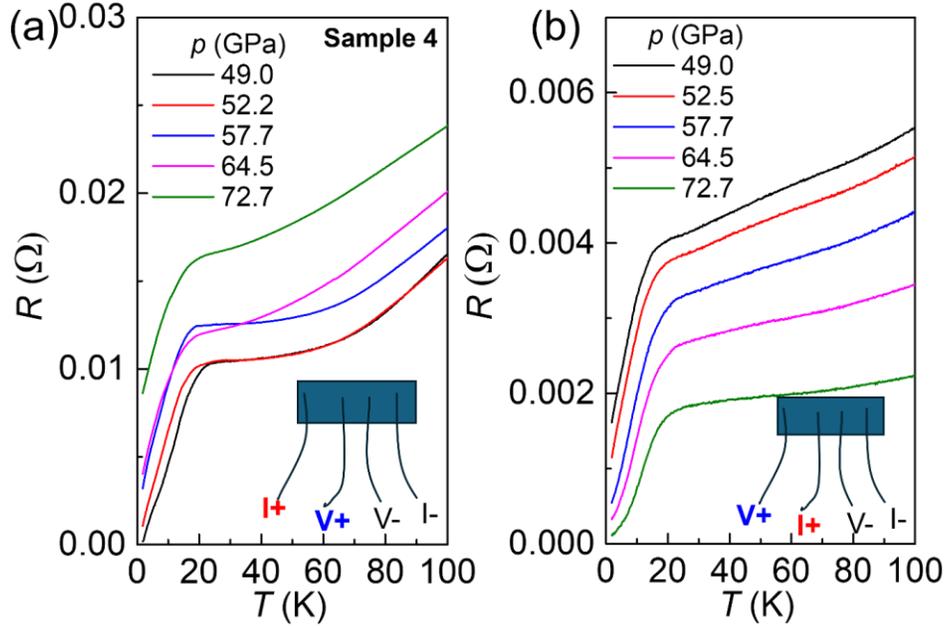

**Figure S3:** (a) $R(T)$ curves at pressures $\geqslant$ 49.0 GPa, obtained using the standard 4-point measurement method. These results are identical to those shown in Fig. 3(d). (b) Measurements at the same pressures as in (a), but with the V+ and I+ channels switched. The insets in (a) and (b) illustrate the schematic configurations of the 4-point measurement setup. These measurements were performed with the Nujol PTM.

## IV. Trapped Flux Measurement

Whereas a weak diamagnetic transition is observed at ~30 K at and above 16 GPa, no clear flux trapping was detected below $T_c$ as shown in Fig. S5. This suggests a filamentary or non-bulk nature of the superconductivity, which lacks a sufficiently large superconducting region for vortices to effectively pin the flux. For comparison, the reader can consider the evolution of the trapped flux signal in CaKFe$_4$As$_4$ [1]. It is noteworthy that for the background subtraction in the trapped flux measurement, we used the 5 GPa results as the background by assuming that the sample does not exhibit superconductivity at 5 GPa.

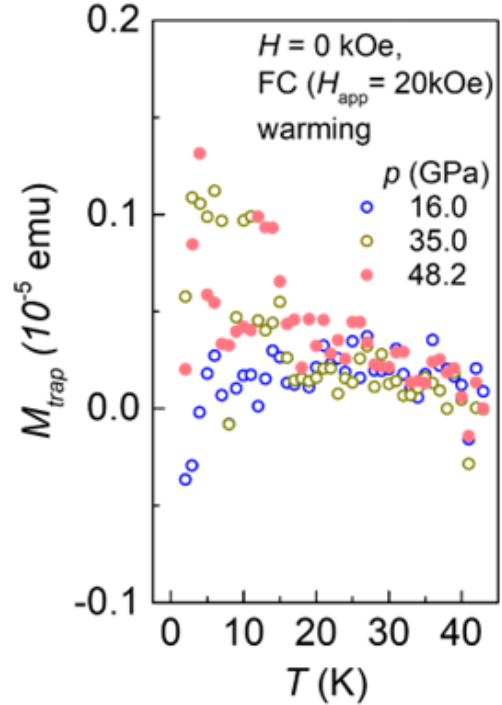

**Figure S5**: Trapped flux measurement, $M_{\text{trap}}(T)$, for the same sample. The long scan results taken at 5 GPa were used as the background for trap flux measurement.

[1] Shuyuan Huyan, Nestor Haberkorn, Mingyu Xu, Paul C. Canfield, Sergey L. Bud'ko, "Competition between the modification of intrinsic superconducting properties and the pinning landscape under external pressure in CaKFe$_4$As$_4$ single crystals" arXiv:2409.03809 (2024).